\documentclass[sigplan,screen,nonacm]{acmart}

\setcopyright{none}

\usepackage[ruled,vlined]{algorithm2e}
\usepackage{graphicx}
\usepackage{textcomp}
\usepackage{xcolor}
\usepackage{listings}
\usepackage{array}
\usepackage{varwidth}
\usepackage{subcaption}

\newcommand{\vectwod}[2]{\ensuremath{\left(\begin{matrix} {#1} \\ {#2}\end{matrix}\right)}}

\def\BibTeX{{\rm B\kern-.05em{\sc i\kern-.025em b}\kern-.08em
    T\kern-.1667em\lower.7ex\hbox{E}\kern-.125emX}}

\begin{document}

\title{Maximal Atomic irRedundant Sets: a Usage-based Dataflow Partitioning 
Algorithm}

\author{Corentin Ferry}
\affiliation{
  \institution{Univ Rennes, CNRS, Inria, IRISA}
  \city{Rennes}
  \country{France}}
\email{cferry@mail.colostate.edu}

\author{Steven Derrien}
\affiliation{
  \institution{Univ Rennes, CNRS, Inria, IRISA}
  \city{Rennes}
  \country{France}}

\author{Sanjay Rajopadhye}
\affiliation{
  \institution{Colorado State University}
  \city{Fort Collins, CO}
  \country{USA}}

\renewcommand{\shortauthors}{Ferry et al.}

\begin{abstract}
Programs admitting a polyhedral representation can be transformed in many ways
for locality and parallelism, notably loop tiling. Data flow analysis can
then compute dependence relations between iterations and between tiles. 
When tiling is applied, certain iteration-wise dependences cross tile 
boundaries, creating the need for inter-tile data communication. Previous work 
\cite{Bondhugula_2013,Dathathri_2013} computes it as the flow-in and flow-out 
sets of iteration tiles. 

In this paper, we propose a partitioning of the flow-out of a tile into the 
maximal sets of iterations that are entirely consumed and incur no redundant 
storage or transfer. The computation is described as an algorithm and 
performed on a selection of polyhedral programs.   We then suggest possible 
applications of this decomposition in compression and memory allocation.
\end{abstract}


\maketitle

\section{Introduction}

A historical and core usage of computers is the acceleration of computations.
The continued demand for precision and speed in signal and data processing 
algorithms has prompted performance engineers to develop carefully tuned 
programs for platforms like graphics processors (GPUs), and even 
domain-specific hardware accelerators (e.g., FPGAs). 

The growth of computation volume and complexity pose multiple challenges to
application developers: because the computing platforms are massively 
parallel at multiple levels (nodes, cores, threads, vectors), they must extract 
just enough parallelism from their programs to use all of what parallelism the 
platform provides, while ensuring without over-constraining the memory hierarchy
in bandwidth and capacity.

In practice, the parallelism actually used is often limited by the data movement 
between CPU cores, GPU threads, nodes, as the data goes through complex memory
hierarchies that introduce latency, resulting in processor stalls waiting for
data. 

Program optimization techniques tend to be made available to developers as 
automatic tools. Certain classes of programs can be analyzed and transformed 
automatically using \emph{polyhedral} analysis and
transformations~\cite{sanjay-fst-tcs, sanjay-thesis, sanjay-dc, quinton-jvsp89,
Feautrier_1991, feautrier92a, feautrier92b}. 

Polyhedral analysis is used to transform programs to improve both parallelism 
and memory access locality (Bondhugula et al., 2008 
\cite{Bondhugula_2008_PLDI, Bondhugula_2008_CC}). Beyond these transformations,
prior work uses  polyhedral analyses to reduce the volume of communicated data
(Bondhugula, Dathathri et al.~\cite{Bondhugula_2013,Dathathri_2013}).

In this paper, we seek to devise optimal sets of data that are communicated 
between tiles of a polyhedral program, with a strict condition to not allow
\emph{redundancy}, both in terms of write (no data is written more than once
into memory) and read (no unused data is read from memory).

This paper is organized as follows: Section \ref{sec:background} introduces the 
concepts used to construct our sets; Section \ref{sec:related-work} gives a
view of other work related to data movement; Section \ref{sec:contribution}
describes the MARS sets and gives their construction 
procedure; Section \ref{sec:analysis} provides examples of constructed sets and
analyses them; Section \ref{sec:applications} gives possible applications of
our work.

\section{Background}
\label{sec:background}

Our work relies on a large stack of polyhedral techniques, ranging from 
analysers to schedulers, and in particular relies on loop tiling. This section
gives a quick glance at the most important techniques we rely on. 

\subsection{Polyhedral model}
The control flow of part of a program is defined by control structures (e.g., 
loops and guards) which conditions, or bounds, can be affine. If this is the 
case, then this control flow admits a \emph{polyhedral representation}: it is
possible to semantically equivalently represent these loops by an 
\emph{iteration space} (the set of all iterations described by the loops and
guards) and a \emph{schedule}, respectively as a set of integer points with 
affine bounds, and an affine map from the iteration space to a multi-dimensional
schedule space.

Likewise, memory accesses performed inside the loop nest which access function
is an affine function of the surrounding iterators can be represented as affine
maps from the iteration space to a collection of \emph{data spaces}.

A \emph{polyhedral model} of an imperative program comprises at least the iteration space,
schedule and memory access information.

\subsection{Dataflow analysis}
A polyhedral model of an imperative program bears read and write access functions, from
which it is possible to compute an iteration-to-iteration dependence pattern.
Dataflow analysis algorithms~\cite{Feautrier_1991} compute this pattern.

The output of dataflow analysis is a dependence graph, which can be expressed
as a \emph{polyhedral reduced dependence graph} (PRDG).  

\subsection{Loop tiling}
In high-performance computing applications, data spaces are too large to fit in
a single level of local memory or cache. Therefore, the iteration space is 
transformed into similarly-shaped, atomic blocks such that the memory footprint
of each block fits in a certain level of cache or local memory. This operation
is called \emph{loop tiling} \cite{Irigoin_1988,Wolf_1991}. 

Formally, loop tiling consists in finding a family of hyperplanes 
$H_i$ such that all dependences that cross the hyperplane (i.e. have
a non-null scalar product with the normal vector of the hyperplane) cross it
in the same direction, which means that all scalar products of dependences 
against the normal vector of the hyperplane must have the same sign.

\emph{In the rest of this paper, we assume the reader is familiar with loop
tiling, and we assume some tiling has been applied to all the programs 
considered. We will then focus on inter-tile communications.}

\section{Related Work}
\label{sec:related-work}

The flow-in and flow-out sets have been extensively studied along with a good 
amount of breakup scenarios by Datharthri et al. (2013) \cite{Dathathri_2013} 
and Bondhugula (2013) \cite{Bondhugula_2013}. We are focusing on one special 
case along the lines of the work of Datharthri et al., with a constraint that 
no point may belong to two communication sets at the same time.

A decomposition of the communicated sets of data may be used for inter-node message
passing in MPI to reduce the amount of traffic. Zhao et al. \cite{Zhao_2021} 
perform a decomposition of the data space of stencils into coarse blocks 
such that fetch and write operations of each block are contiguous,
and blocks are laid out according to the consuming neighbors so that a series of
blocks is retrieved in one contiguous message. A supporting graph data structure
provides the addresses for each of the blocks. This work seeks an optimal memory 
layout in terms of number of communications, and does so
without the flow-in and flow-out sets or a polyhedral representation. Our work
generalizes the idea using the polyhedral framework.

\section{MARS: Maximal Atomic irRedundant Sets}
\label{sec:contribution}

This section presents the Maximal Atomic irRedundant Sets (MARS). It is laid 
out as follows: first, we give a definition of MARS and properties that they 
match. Then, we introduce an algorithm to construct these sets along with
an example.

\subsection{Notations and hypotheses}

To compute the MARS, we need a program with a polyhedral model, to which the
tiling transformation is legal along given hyperplanes. We restrict ourselves
to the case where the dependences are uniform, and therefore we can consider
individual \emph{dependence vectors}. The uniformity of the dependence pattern
guarantees that, assuming an infinite iteration space, all tiles of the same 
shape feature the same MARS. Tile sizes are assumed to be constant, but
they can be made runtime parameters using the idea from \cite{sanjay-toplas2012}.

We also assume that each statement writes to a single memory location. 
Therefore, we can interchangeably use an iteration and the value it produces
(data).

Additionally, we will use the following notations:
\begin{itemize}
  \item $\mathcal{D}$ designates the iteration space, of dimension $N$.
  \item There are $T$ tiling hyperplanes, $H_i$ for 
  	$i \in \left\lbrace 1,\dots,T \right\rbrace$.
  \item There are $D$ dependence vectors, $\vec{b}_j$ for 
  	$j \in \left\lbrace 1,\dots,D \right\rbrace$.
  	The PRDG is the set of all dependence vectors.
  \item The \emph{non-trivial parts} of a set $E$ are all the non-empty subsets 
  of $E$. It is noted $\mathcal{P}_n(E)$.
  For instance: $$ \mathcal{P}_n\left(\lbrace 1, 2, 3 \rbrace\right) = 
	\left\lbrace \lbrace 1 \rbrace, \lbrace 2 \rbrace, \lbrace 3 \rbrace,
	\lbrace 1,2 \rbrace, \lbrace 1,3 \rbrace, \lbrace 2,3 \rbrace, 
	\lbrace 1,2,3 \rbrace \right\rbrace$$
  \item The modulo operator is noted $a \text{ mod } b$ and congruences are 
  noted $a \equiv r [b]$.
\end{itemize}

\subsection{Definition}

Maximal Atomic irRedundant Sets (MARS) are defined as the maximal sets of 
iterations that satisfy the following property:

\begin{center}
\textbf{All-consumed per tile} (ACT): if a tile consumes data produced by an 
iteration $\vec{x}$ inside a MARS, then it consumes all the data produced by the 
same tile as $\vec{x}$ inside that MARS.
\end{center}

\begin{figure}
    \centering
    \subcaptionbox{a}{\includegraphics[width=0.45\columnwidth]
    	{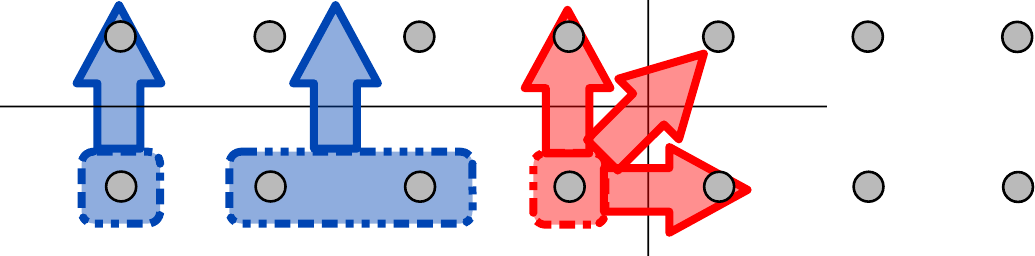}}\hfill
    \subcaptionbox{b}{\includegraphics[width=0.45\columnwidth]
    	{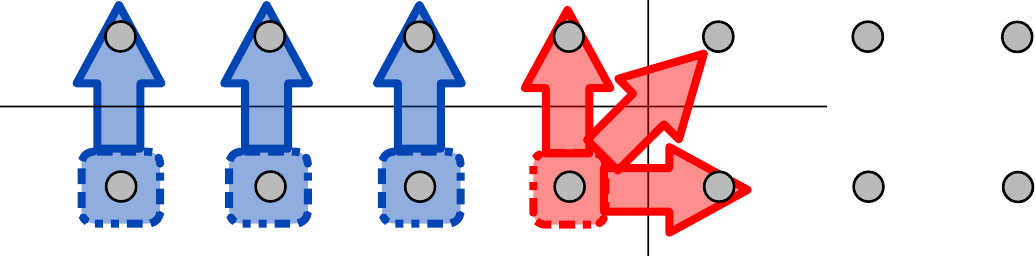}}
    \caption{Sets of iterations for Smith-Waterman matching (ACT) that are not 
    MARS}
    \label{fig:mars-unique}
\end{figure}

Figure \ref{fig:mars-unique} shows this property is not sufficient for 
the sets to be unique. However, the maximal such sets are unique.

\subsection{Computation}
The all-consumed property stated above is equivalent to saying that all
data inside a MARS is consumed by exactly the same tiles. Therefore,
if two distinct tiles consume a MARS $M$, then given the all-consumed property, 
both consumer tiles use all of the points of $M$.

We propose a construction by breaking up the \emph{flow-out set} of each tile. 
There is an equivalent construction with the \emph{flow-in set} of each tile, 
and both constructions lead to the maximal sets that repect the (ACT) property. 

\subsubsection{Flow-out set}

\begin{figure}
    \centering
    \includegraphics[width=\columnwidth]{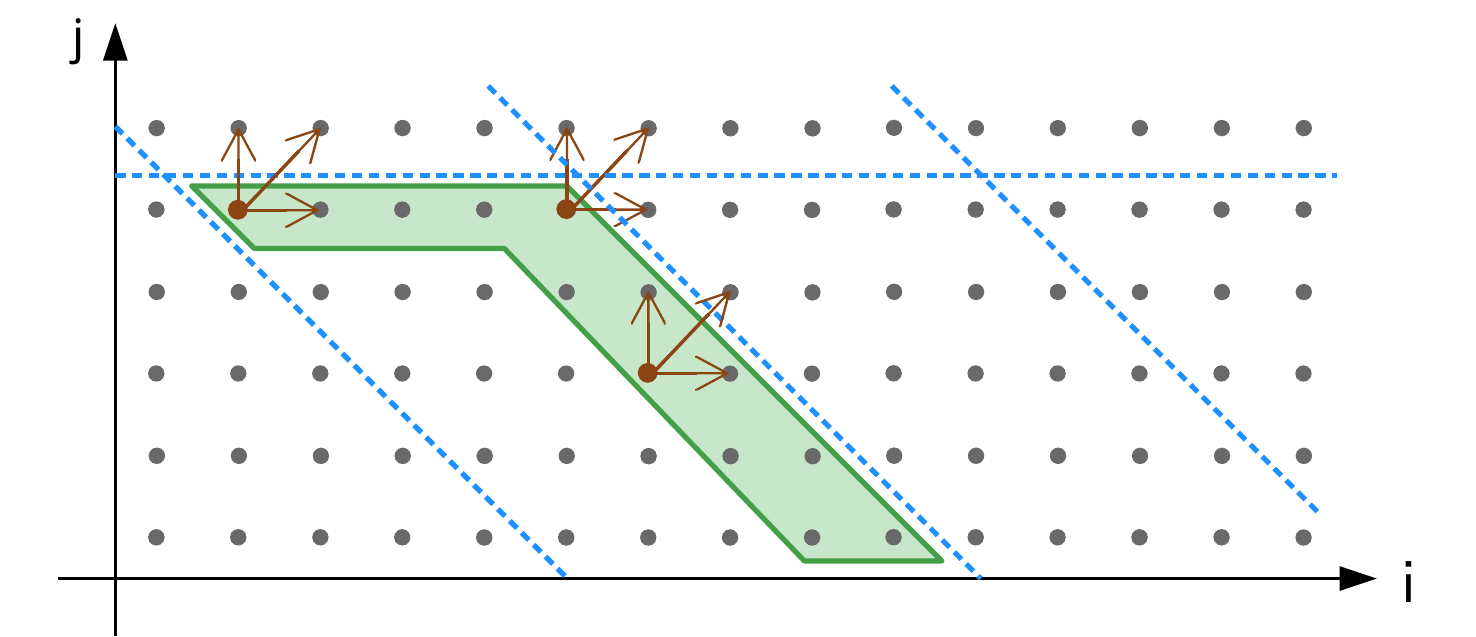}
    \caption{Flow-out set for a skewed tile of a Smith-Waterman kernel.
    The arrows correspond to the dependence pattern (PRDG).}
    \label{fig:mars-running-consumed-points}
\end{figure}

We start by introducing the flow-out set, with a different view than
\cite{Bondhugula_2013}. The flow-out set of a tile is defined as all the 
iterations which have at least one consumer iteration outside the tile. An 
example is given in Figure \ref{fig:mars-running-consumed-points} for a 
Smith-Waterman kernel with skewed tiles. Notably, we see that despite the
dependences being all unit or null along each axis, the ``thickness'' of the
flow-out may be greater than one (in this case, with the diagonal dependence). 

The tile-wise flow-out can be expressed as follows: given a tile $S$,
$$ \varphi_O(S) = \left\lbrace \vec{x} \in S : \vec{x} + \vec{b}_1 \in 
\mathcal{D} \backslash S \vee \dots \vee \vec{x} + \vec{b}_D \in 
\mathcal{D} \backslash S \right\rbrace $$

However, this formulation is missing information on the tile the dependence
vectors lead to. We therefore introduce a finer formulation with the individual
contribution of each dependence that traverses a tiling hyperplane, i.e. that
crosses tile boundaries. This only requires knowledge of the PRDG (dependence
vectors) alongside the tiling hyperplanes and the domain, and can be done as 
in Algorithm~\ref{alg:flow-out}.

\begin{algorithm}
\SetAlgoLined
\KwIn{$\begin{aligned}
\mathcal{D} &= \text{iteration space}, \\
B = \left\lbrace \vec{b}_i : i=1,\dots,D\right\rbrace &= \text{PRDG}, \\ 
\mathcal{H} = \left\lbrace H_i  : i=1,\dots,T \right\rbrace 
	&= \text{tiling hyperplanes}
\end{aligned}$}
\KwResult{$\varphi_O$ = Flow-out set}
\For{$H \in \mathcal{H}$}{
	$\vec{c} = (c_i)_{i=1, \dots, N} \text{ normal vector to } H$\;
	$s = \text{ tile size for hyperplane } H$\;
	\For{$\vec{b} \in B$}{
		\tcp{Flow-out iterations for dependence $\vec{b}$
			crossing hyperplane $H$}
		$m = \vec{c} \cdot \vec{b}$\;
		$\begin{aligned} F_{H, \vec{b}} = & \left\lbrace \vec{x} = (x_i)_{i=1, \dots, N} \in \mathcal{D} : \right.\\ 
			& \left. -m \leqslant {\textstyle\sum_i} (c_i x_i) \text{ mod } s < 0 \wedge \vec{x} + \vec{b} \in \mathcal{D} \right\rbrace\end{aligned}$\;
	}
}
$\varphi_O = \bigcup_{H \in \mathcal{H}} 
	\bigcup_{\vec{b} \in B} F_{H, \vec{b}} $\;
\Return{$\varphi_O$}
\caption{Computing the the flow-out set using contributions from each dependence}
\label{alg:flow-out}
\end{algorithm}

Following Algorithm \ref{alg:flow-out}, the flow-out is then the union of 
the contributions of all dependences to it:
$$ \varphi_O = \bigcup_{H \in \mathcal{H}} 
	\bigcup_{\vec{b} \in B} F_{H, \vec{b}} $$

and then the flow-out of a given tile is given by intersecting the domain of
a tile to the flow-out iterations of the entire domain: 
$$\varphi_O(S) = \varphi_O \cap S$$

\subsubsection{MARS partitioning of the flow-out set}

We explain how our partitioning scheme is done in this subsection.

\begin{figure}
    \centering
    \includegraphics[width=0.75\columnwidth]{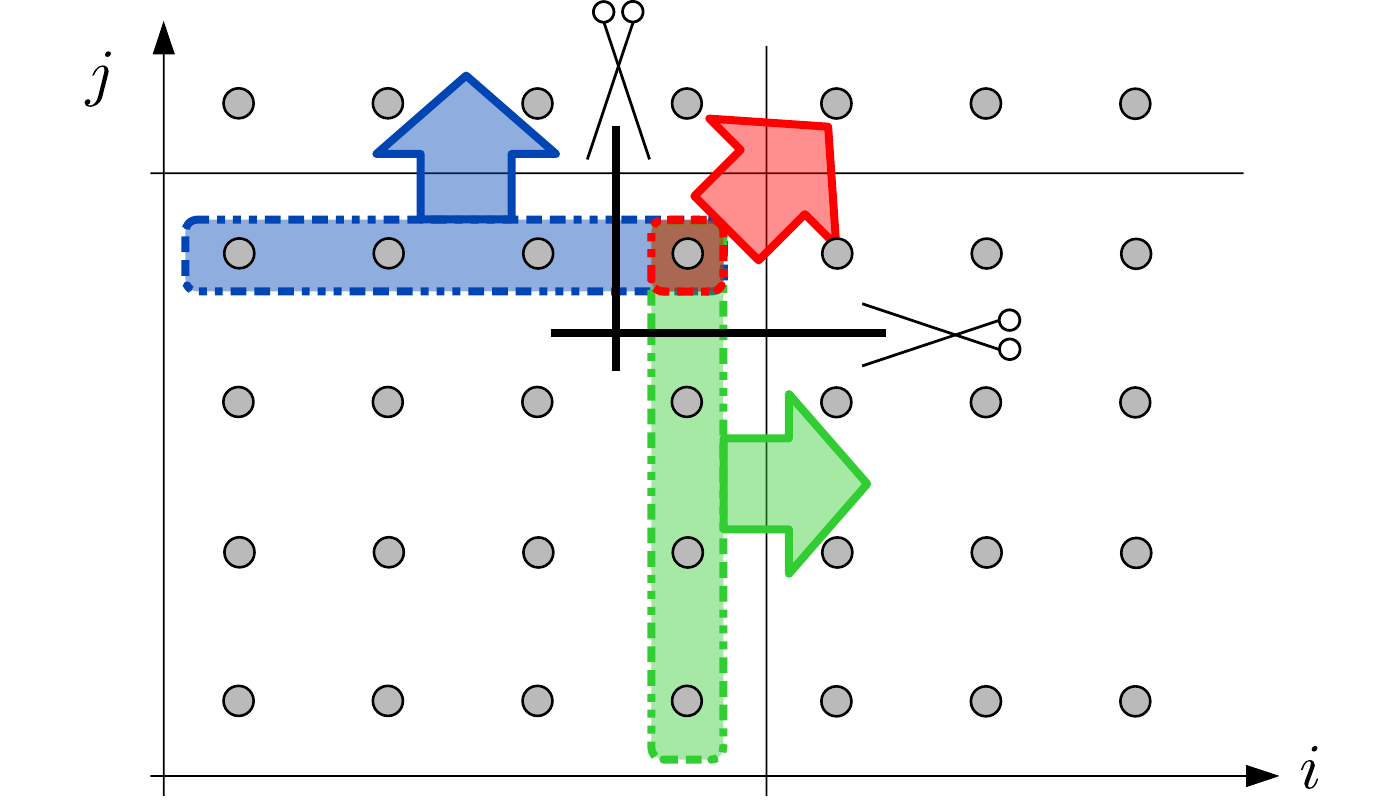}
    \caption{Flow-out set of a tile intersected with each consumers tile's 
    flow-in set. The breakup we propose splits iterations that have different
    consumer tiles.}
    \label{fig:mars-split}
\end{figure}

\paragraph{Principle}
The flow-out set can be partitioned into MARS in such a way that, for any given
tile, all iterations it produces in a MARS have the exact same consumer tiles. 
The partitioning idea is illustrated in Figure~\ref{fig:mars-split}.

The partitioning is done by computing those subsets of the flow-out for which,
given select tiling hyperplanes, any dependence crosses all these tiling 
hyperplanes, and no dependence crosses any other tiling hyperplane.

We browse all possible consumers by applying the above on all combinations of 
tiling hyperplanes. As there are $T$ tiling hyperplanes, there are $2^T-1$ 
possible consumer tiles, and therefore at most $2^{2^T-1}-1$ MARS. 

\paragraph{Example}
We can construct the MARS for a Smith-Waterman kernel, which has the 
following characteristics:

\begin{itemize}
  \item Domain: $\mathcal{D} = \left\lbrace \vectwod{i}{j} : (i, j) \in [0,100]^2 
  	\right\rbrace$
  \item PRDG: $B = \left\lbrace 
  \vec{b}_1 = \vectwod{1}{0},
  \vec{b}_2 = \vectwod{0}{1},
  \vec{b}_3 = \vectwod{1}{1}
  \right\rbrace$ ($D=3$)
  \item Tiling hyperplane families: \\
  $\mathcal{H} = \left\lbrace 
  \left\lbrace \mathcal{H}_1 : i \equiv 0 [4] \right\rbrace, 
  \left\lbrace \mathcal{H}_2 : j \equiv 0 [4] \right\rbrace \right\rbrace $ ($T=2$)
  \item Normal vectors: $\left\lbrace 
  \vec{c}_1 = \vectwod{1}{0},
  \vec{c}_2 = \vectwod{0}{1}
  \right\rbrace$
  \item Tile sizes: $\left\lbrace 4, 4 \right\rbrace$
\end{itemize}  

We can first notice that dependence $\vec{b}_1$ does not cross hyperplane 
$\mathcal{H}_2$, and likewise dependence $\vec{b}_2$ does not cross hyperplane 
$\mathcal{H}_1$.

Let's start by considering those dependences that cross any hyperplane of the 
$\mathcal{H}_1$ family and none of $\mathcal{H}_2$. 

To have a dependence cross hyperplane $\mathcal{H}_1$, the source iteration 
$\vec{x} = \vectwod{i}{j}$ must be such that, 
if $\vec{b} = \vectwod{b_i}{b_j}$, then 
$(i \text{ mod } 4) + b_i \geqslant 4$ or $(i \text{ mod } 4) + b_i < 0$. 
Because all $\vec{b}$s only have positive coordinates, let's only consider the 
case $(i \text{ mod } 4) + b_i \geqslant 4$. 

From dependence $\vec{b}_1$, we get $i \text{ mod } 4 \geqslant 3$; from dependence 
$\vec{b}_3$, we also get $i \text{ mod } 4 \geqslant 3$. Therefore, the set of 
points such that \emph{any} dependence crosses $\mathcal{H}_1$ is:
$$ \left\lbrace \vec{x} = \left(\begin{matrix} i \\ j\end{matrix}\right) \in 
	\mathcal{D} : i \text{ mod } 4 \geqslant 3 \right\rbrace$$
or equivalently
$$ \left\lbrace \vec{x} = \left(\begin{matrix} i \\ j\end{matrix}\right) \in 
	\mathcal{D} : i \equiv 3 [4]\right\rbrace$$

We compute the subset of these points for which $\mathcal{H}_2$ is crossed. The 
condition to cross $\mathcal{H}_2$ is that, if $\left(\begin{matrix} i \\ 
j\end{matrix} \right) = (\vec{x} + \vec{b})$, then 
$j \text{ mod } 4 + b_j \geqslant 4$, which means $j \text{ mod } 4 
\geqslant 3$ with both dependences $\vec{b}_1$ and $\vec{b}_2$. 
We therefore get that the points from which $\mathcal{H}_1$ is crossed and not 
$\mathcal{H}_2$ is:
$$ \left\lbrace \vec{x} = \left(\begin{matrix} i \\ j\end{matrix}\right) \in 
	\mathcal{D} : i \equiv 3 [4] \wedge \neg(j \equiv 3 [4]) \right\rbrace$$

We can do the same procedure to cross only $\mathcal{H}_2$, and both of
$\mathcal{H}_1$ and $\mathcal{H}_2$, which yield the following sets:

$$ \left\lbrace \vec{x} = \left(\begin{matrix} i \\ j\end{matrix}\right) \in 
	\mathcal{D} : \neg(i \equiv 3 [4]) \wedge (j \equiv 3 [4]) \right\rbrace $$
and 
$$ \left\lbrace \vec{x} = \left(\begin{matrix} i \\ j\end{matrix}\right) \in 
	\mathcal{D} : i \equiv 3 [4] \wedge j \equiv 3 [4] \right\rbrace$$

\begin{figure}
    \centering
    \includegraphics[width=0.75\columnwidth]{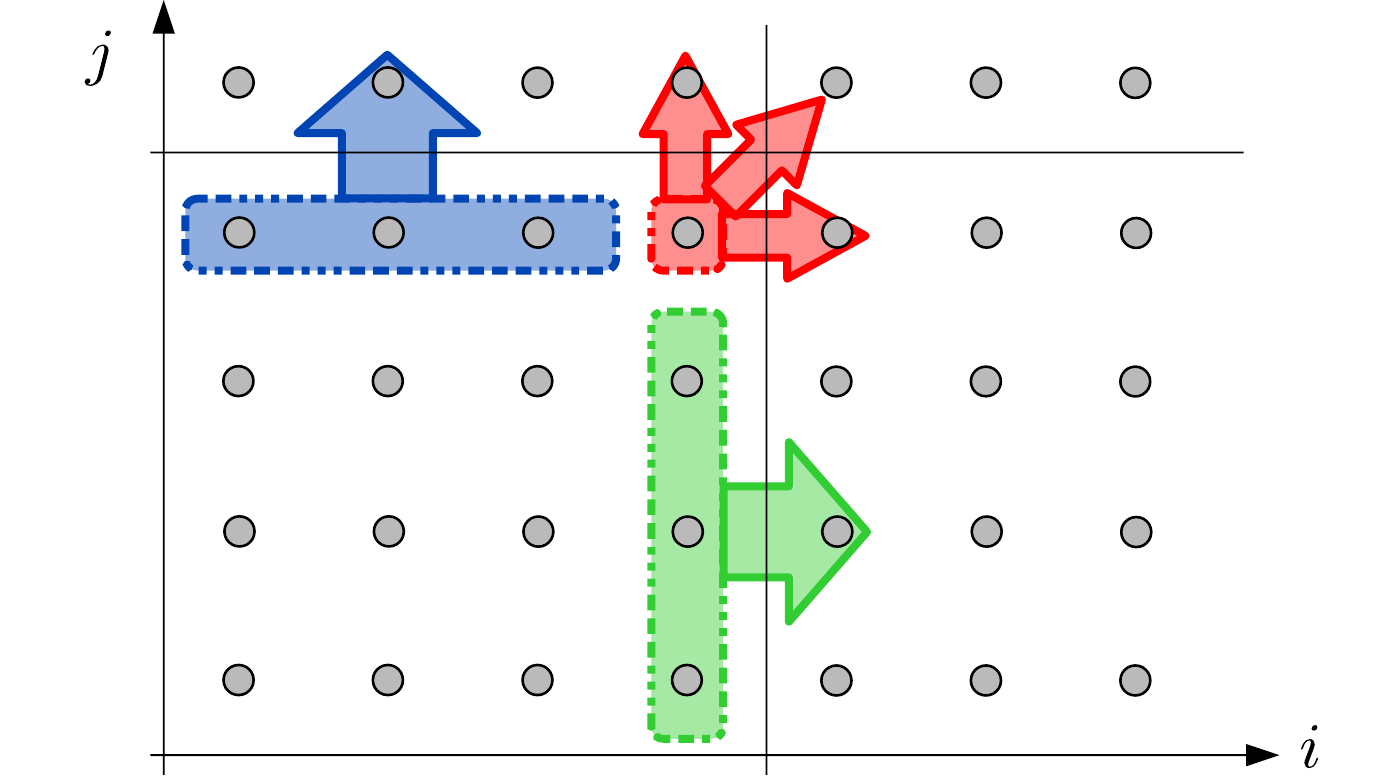}
    \caption{MARS and their consumers for Smith-Waterman using square tiling.}
    \label{fig:mars-sw2d-square}
\end{figure}

Those sets are the MARS we were looking for, and correspond to those in 
Figure~\ref{fig:mars-sw2d-square}.

\paragraph{Algorithm}

Algorithm \ref{alg:mars} gives the computation procedure to construct all MARS
for all tiles. 

In this algorithm, crossing a hyperplane is a shortcut for the property used in 
Algorithm~\ref{alg:flow-out}. Assume $\vec{c} = (c_i)$ is the normal vector
to a hyperplane $H$, $s$ is the tile size along that hyperplane, 
$\vec{b}$ is a dependence vector, and $\vec{x} = (x_i) \in \mathcal{D}$. 
Let $m = \vec{c}\cdot\vec{b}$ assuming $m > 0$. Then:
$$ \vec{x} + \vec{b} \text{ crosses }H \Leftrightarrow 
-m \leqslant {\textstyle\sum_i} (c_i x_i) \text{ mod } s < 0$$

\begin{algorithm}
\SetAlgoLined
\KwIn{$\begin{aligned}
\mathcal{D} &= \text{iteration space}, \\
B = \left\lbrace \vec{b}_i : i=1,\dots,D\right\rbrace &= \text{PRDG}, \\ 
\mathcal{H} = \left\lbrace H_i  : i=1,\dots,T \right\rbrace 
	&= \text{tiling hyperplanes}
\end{aligned}$}
\KwResult{$\mathcal{M}$ = partition of flow-out into MARS}
$\mathcal{M} = \varnothing$\;
$\mathcal{T} = \mathcal{P}_n\left(\mathcal{H}\right)$\tcc*{All neighboring tiles}
\For{$I \in \mathcal{P}_n\left(\mathcal{T}\right)$}{
	$E = \mathcal{T} \backslash I$\;
	\tcp{$A$: all tiles in $I$ must be reached by $\geqslant 1$ dependence}
	\For{$T \in I$} {
		$N = \mathcal{H} \backslash T$\;
		\For{$\vec{b} \in B$} {
			\tcp{$P_{T,\vec{b}}$: $\vec{b}$ crosses all hyperplanes of $T$ 
			and no others}
			$\begin{aligned} P_{T,\vec{b}} = & {\textstyle \left\lbrace \vec{x} \in \mathcal{D} : 
				\bigwedge_{H \in T}\left(\vec{x} + \vec{b} \text{ crosses } H
				\right) \wedge\right.} \\
			    &  {\textstyle \left. 
				\bigwedge_{H \in N}\neg\left(\vec{x} + \vec{b} \text{ crosses } H
				\right)	\wedge \vec{x} + \vec{b} \in \mathcal{D} \right\rbrace }\end{aligned}$\;
		}
	}
	$A = \bigcap_{T \in I}\bigcup_{\vec{b} \in B} P_{T, \vec{b}}$\;
	\tcp{$S$: no tiles in $E$ may be reached by any dependence}
	\For{$\vec{b} \in B$} {
		\For{$T \in E$}{
			$N = \mathcal{H} \backslash T$\;
			\tcp{$Q_{T,\vec{b}}$: $\vec{b}$ crosses an hyp. in $N$ or
			doesn't cross an hyp. in $T$}
			$\begin{aligned} Q_{T,\vec{b}} = & {\textstyle \left\lbrace \vec{x} \in \mathcal{D} :
				\left[ \bigvee_{H \in T}\neg\left(\vec{x} + \vec{b} 
				\text{ crosses } H \right) \vee \right.\right.}\\
                & {\textstyle \left. \left.
				\bigvee_{H \in N}\left(\vec{x} + \vec{b} \text{ crosses } H
				\right)
				\right] \wedge  \vec{x} + \vec{b} \in \mathcal{D}
				\right\rbrace }\end{aligned}$\;
		}
	}
	$S = \bigcap_{T \in E}\bigcap_{\vec{b} \in B} Q_{T, \vec{b}}$\;
	$\mathcal{M} = \mathcal{M} \cup \left\lbrace A\cap S \right\rbrace$\;
}
\Return{$\mathcal{M}$}
\caption{Computing the MARS}
\label{alg:mars}
\end{algorithm}

By intersecting the MARS obtained from Algorithm~\ref{alg:mars} with individual
tiles, we obtain a decomposition of every tile's flow-out set into MARS. They
then hold the following two properties:
\begin{itemize}
  \item Each MARS is composed of iterations from a single tile,
  \item Each MARS is entirely consumed by every of its consumer tiles.
\end{itemize}

\subsection{Dual view: flow-in}
Equivalently to partitioning the flow-out set into MARS, it is possible to 
compute the partitioning of the flow-in set of each tile into MARS. 
The flow-in set is computed with the same algorithm as the flow-out, 
using the opposite of the dependence vectors.

Intersecting the MARS created by Algorithm~\ref{alg:mars} (not broken up into
individual tile-wise MARS) with the obtained tile-wise flow-in set then gives
a breakup into MARS. This partitioning can be used to figure out which MARS
every tile should fetch from other tiles as an input. 

\section{Implementation and Analysis}
\label{sec:analysis}

It is possible to express all MARS using polyhedral tools (ISL) provided the 
tile sizes are constant. However, using the idea from \cite{sanjay-toplas2012}, 
it is possible to use parametric tile sizes by adding those sizes as 
additional parameters. We have implemented a MARS computer in Python using 
ISLPy.

\subsection{Usage of the MARS computer}
A MARS computer is available at \url{https://github.com/cferr/mars.git}.

To compute the MARS for a given program and tiling hyperplanes, it needs input 
that can be computed using publicly available tools:
\begin{itemize}
	\item The polyhedral model of a program, to be extracted for instance 
	using PET \cite{Verdoolaege_2012}; 
	\item Dependence vectors, obtained using array dataflow analysis
	e.g. using \texttt{iscc}; 
	\item Legal tiling hyperplanes, found for instance by calling PLuTo; the
	standard equation and the normal vectors to these hyperplanes are to be 
	provided. 
\end{itemize}

The MARS computer then runs Algorithm~\ref{alg:mars}. A visualization of the 
MARS is given with \texttt{islplot} when the iteration space is two- or 
three-dimensional. For two-dimensional iteration spaces, the MARS in the entire
iteration space can be visualized; for three-dimensional spaces, a sample tile
needs to be provided and the MARS specific to that tile will be shown.

\subsection{Results}
We have run the MARS computer against a series of uniform dependence benchmarks.
This section evaluates the result on the following questions:
\begin{itemize}
  \item How many MARS are there per tile? 
  \item What is the dimensionality of the result MARS?
  In particular, how many singleton MARS are there?
\end{itemize}

\subsubsection{Evaluated applications}
The MARS computer has been used on the following applications:
\begin{itemize}
	\item \texttt{sw}: Smith-Waterman dynamic programming algorithm for sequence
	alignment;
	\item \texttt{jacobi-1d}: Jacobi 1D stencil;
	\item \texttt{canonical-3d}: Artificial 3-dimensional example that has a
	dependence along each canonical axis;
	\item \texttt{gemm}: GEMM (BLAS) implementation from PolyBench 
	\cite{Polybench};
	\item \texttt{seidel-2d}: Seidel 2D stencil implementation from PolyBench;
	\item \texttt{jacobi-2d}: Jacobi 2D stencil implementation from PolyBench;
\end{itemize}

The \texttt{jacobi-2d} benchmark is exploited twice, with different tiling 
schemes: one is rectangular tiling combined with skewing, the other one is
diamond tiling \cite{Bondhugula_2017}. We will refer to them respectively as 
\texttt{jacobi-2d-r} and \texttt{jacobi-2d-d}. 

\begin{table*}
\begin{tabular}{ccV{4cm}V{3cm}ccccc}
Dims & Application & Dependences & Tiling hyperplanes & \# Cons. tiles & Nb MARS & Singletons \\
\hline
2    & \texttt{sw} & $(1, 0), (0, 1), (1, 1)$ & $i + j$, $j$ & 3 & 4 & 2 \\
2    & \texttt{jacobi-1d} & $(1, -1), (1, 0), (1, 1)$ & $t+i$, $t-i$ & 3 & 4 & 2 \\
3    & \texttt{canonical-3d} & $(1, 0, 0), (0, 1, 0), (0, 0, 1)$ & $i$, $j$, $k$ & 3 & 7 & 1 \\
3    & \texttt{gemm} & $(0, 1, 0)$ & $i$, $j$, $k$ & 1 & 1 & 0 \\
3    & \texttt{seidel-2d} & $(0, 1, 1)$, $(0, 0, 1)$, $(1, -1, 1)$, 
							$(0, 1, 0)$, $(1, 0, 0)$, $(1, -1, 0)$, 
							$(0, 1, -1)$, $(1, 0, -1)$, $(1, -1,-1)$ & $t$, $t + i$, $4t + 2i + j$ & 7 & 13 & 2 \\
3    & \texttt{jacobi-2d-r} & $(1, 0, 1)$, $(1, 1, 0)$, $(1, 0, 0)$, 
							$(1, -1, 0)$, $(1, 0, -1)$ & $t$, $t+i$, $t+j$ & 7 & 13 & 4 \\
3    & \texttt{jacobi-2d-d} & $(1, 0, 1)$, $(1, 1, 0)$, $(1, 0, 0)$, 
							$(1, -1, 0)$, $(1, 0, -1)$ & $t+i$, $t+j$, $t-i$, $t-j$ & 15 & 34 (26) & 6 \\
\end{tabular}
\caption{Results obtained from the MARS computer}
\label{tab:results}
\end{table*}

\begin{figure*}
\centering
	\begin{subfigure}{0.4\textwidth}
	\centering
	\includegraphics[height=5cm]{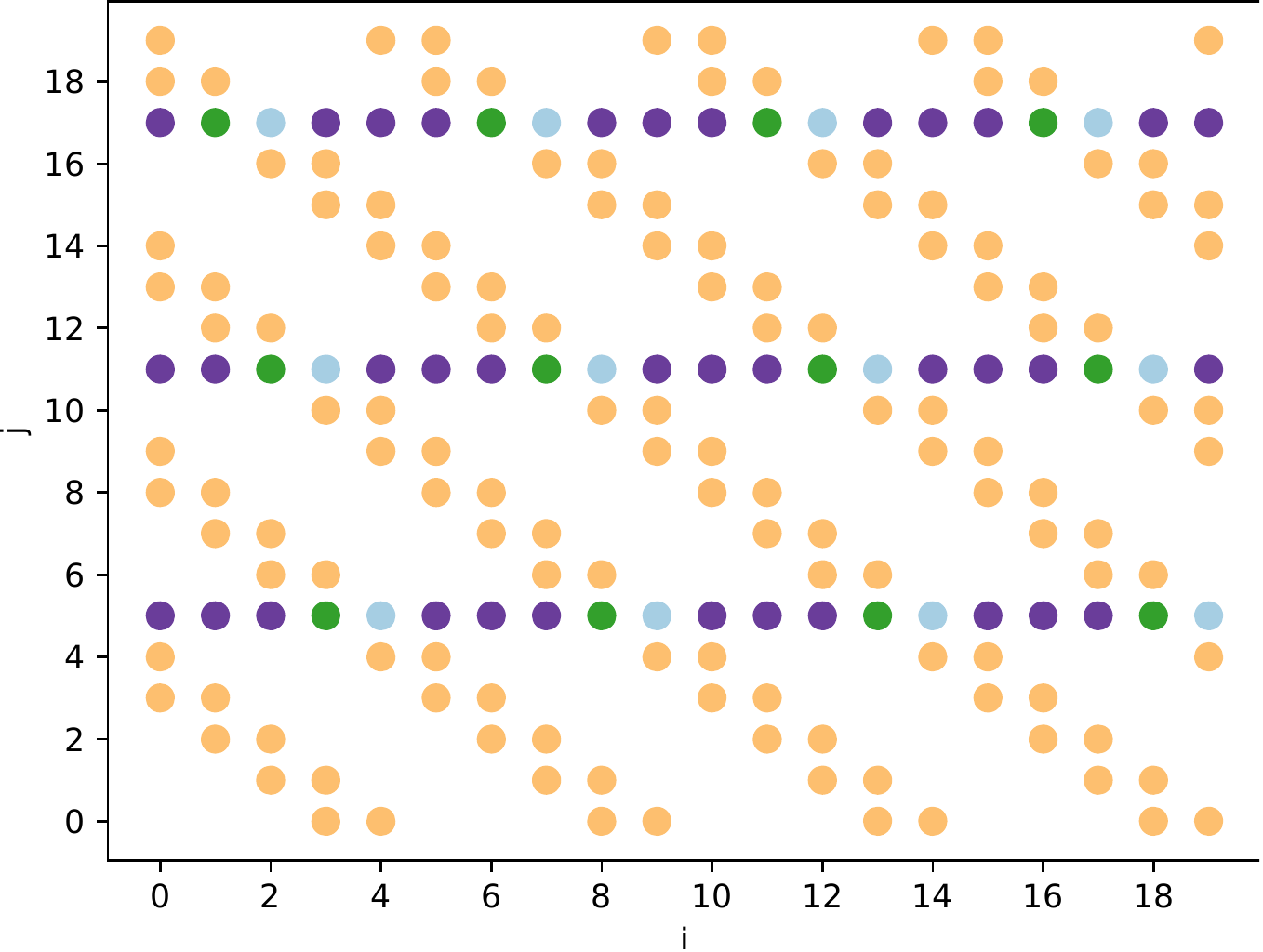}
    \caption{\texttt{sw}}
    \label{fig:mars-sw}
    \end{subfigure}
    \begin{subfigure}{0.4\textwidth}
    \centering
	\includegraphics[height=5cm]{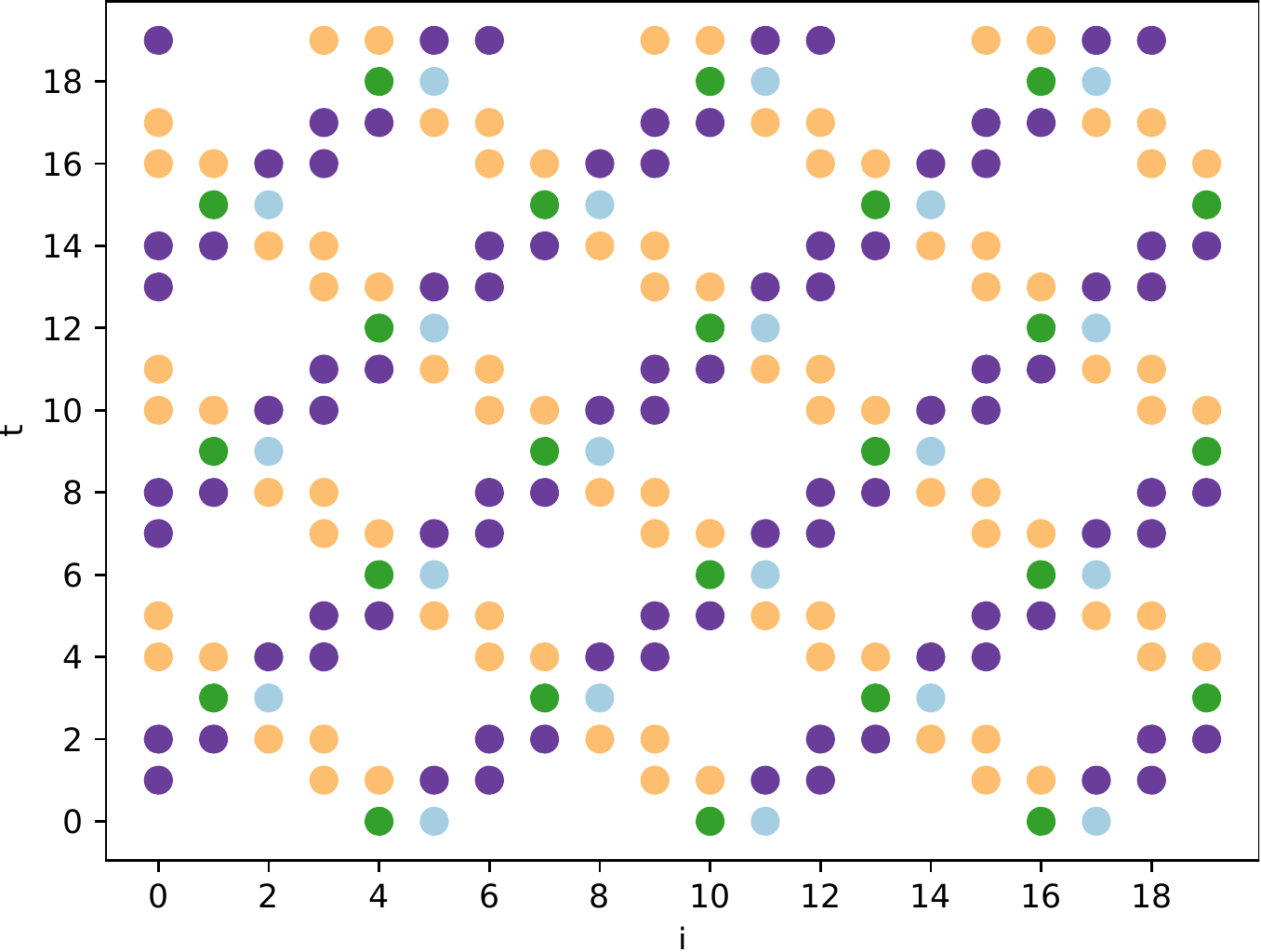}
    \caption{\texttt{jacobi-1d}}
    \label{fig:mars-jacobi-1d}
    \end{subfigure}
    \begin{subfigure}{0.4\textwidth}
    \centering
	\includegraphics[height=5cm]{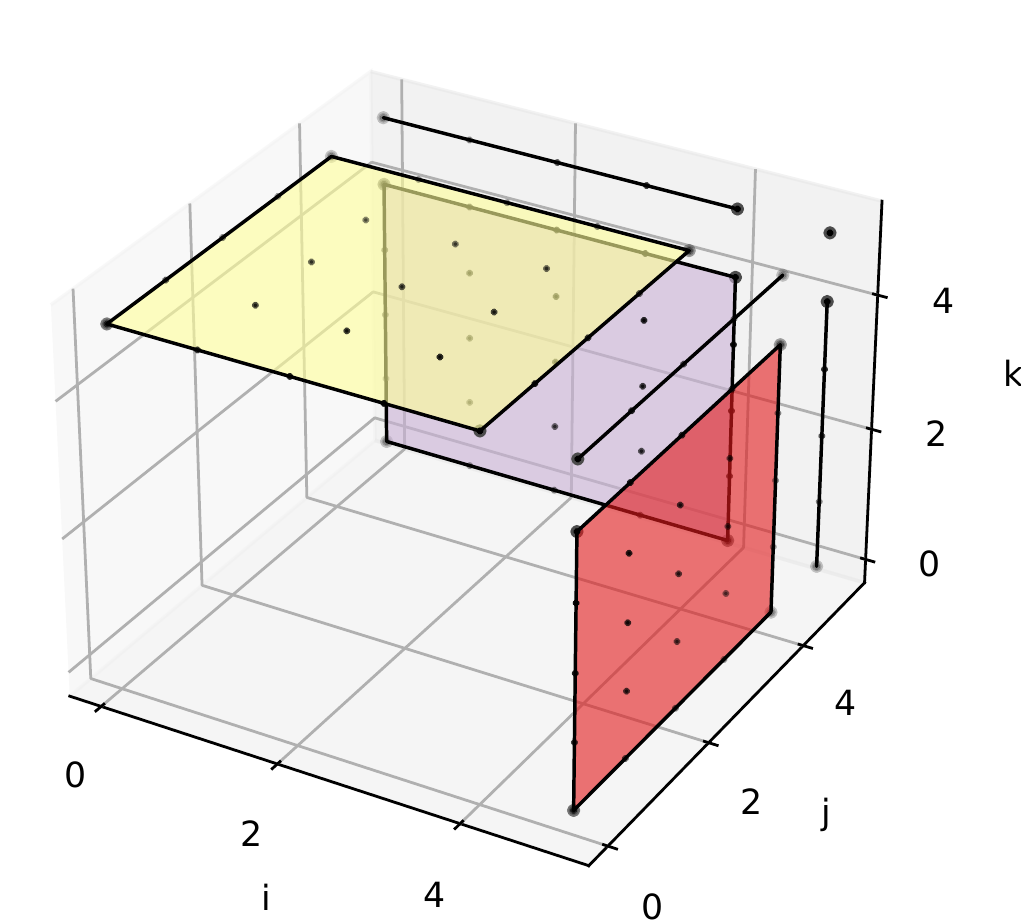}
    \caption{\texttt{canonical-3d}}
    \label{fig:mars-canonical-3d}
    \end{subfigure}
    \begin{subfigure}{0.4\textwidth}
    \centering
	\includegraphics[height=5cm]{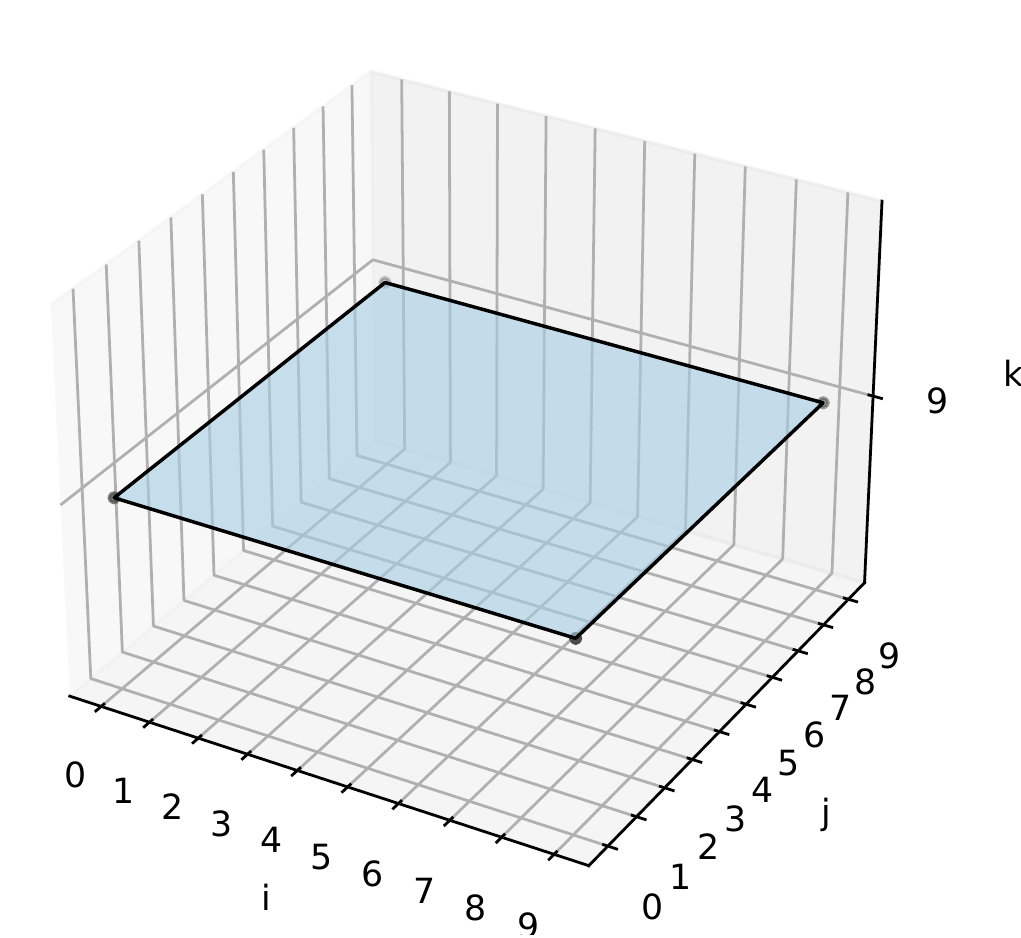}
    \caption{\texttt{gemm}}
    \label{fig:mars-gemm}
    \end{subfigure}
    \begin{subfigure}{0.4\textwidth}
    \centering
	\includegraphics[height=5cm]{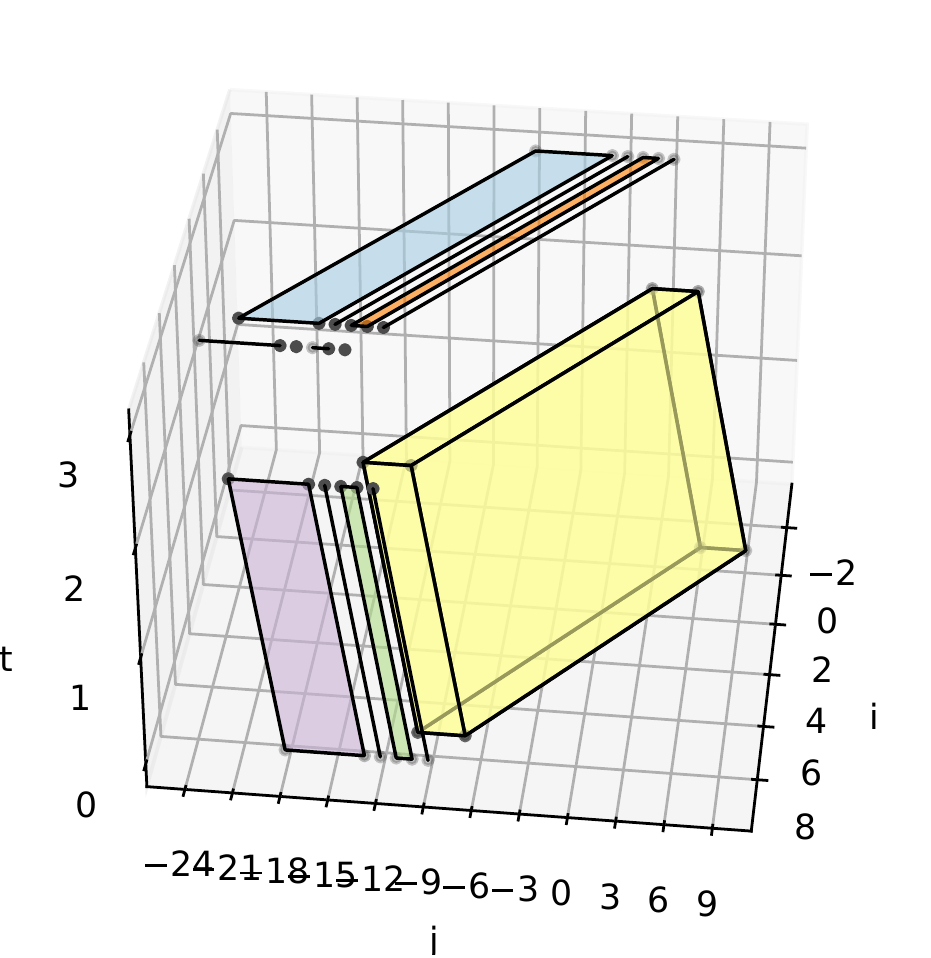}
    \caption{\texttt{seidel-2d}}
    \label{fig:mars-seidel-2d}
    \end{subfigure}
    \begin{subfigure}{0.4\textwidth}
    \centering
	\includegraphics[height=5cm]{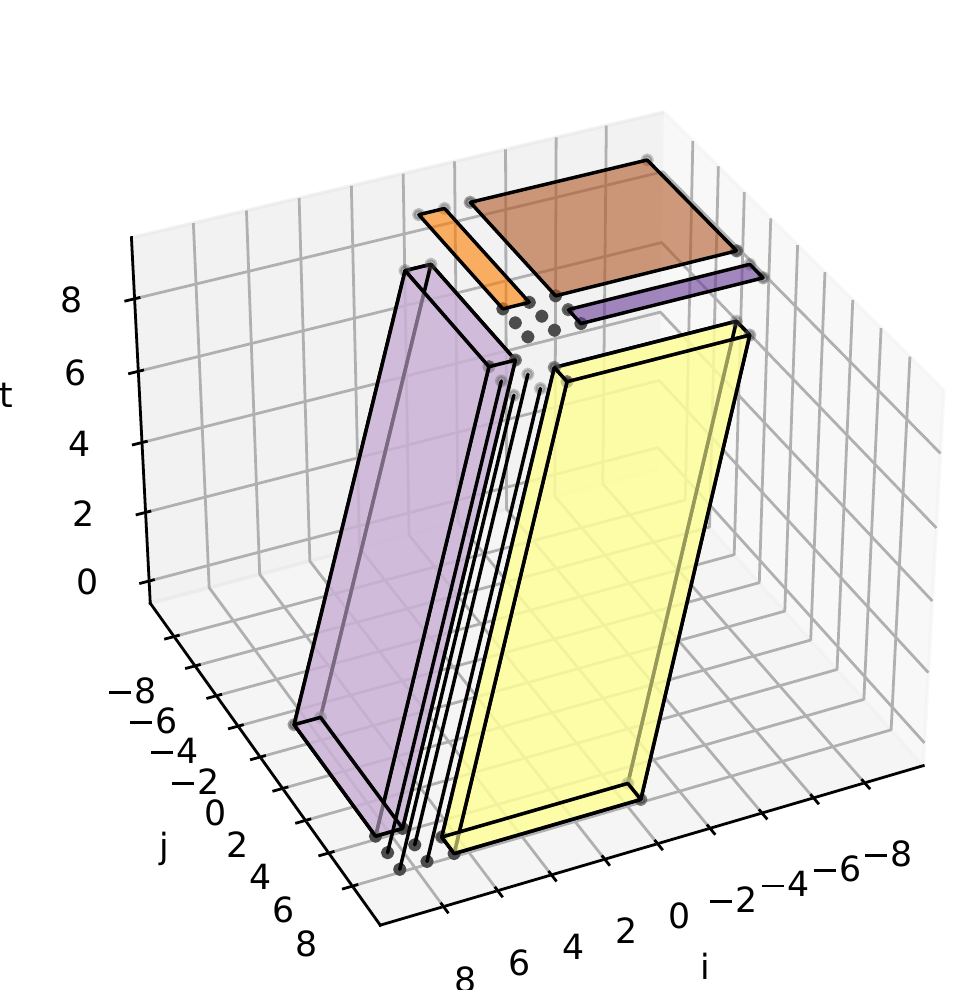}
    \caption{\texttt{jacobi-2d-r} (skewing + rectangular tiling)}
    \label{fig:mars-jacobi-2d-skewed}
    \end{subfigure}
    \begin{subfigure}{0.4\textwidth}
    \centering
	\includegraphics[height=5cm]{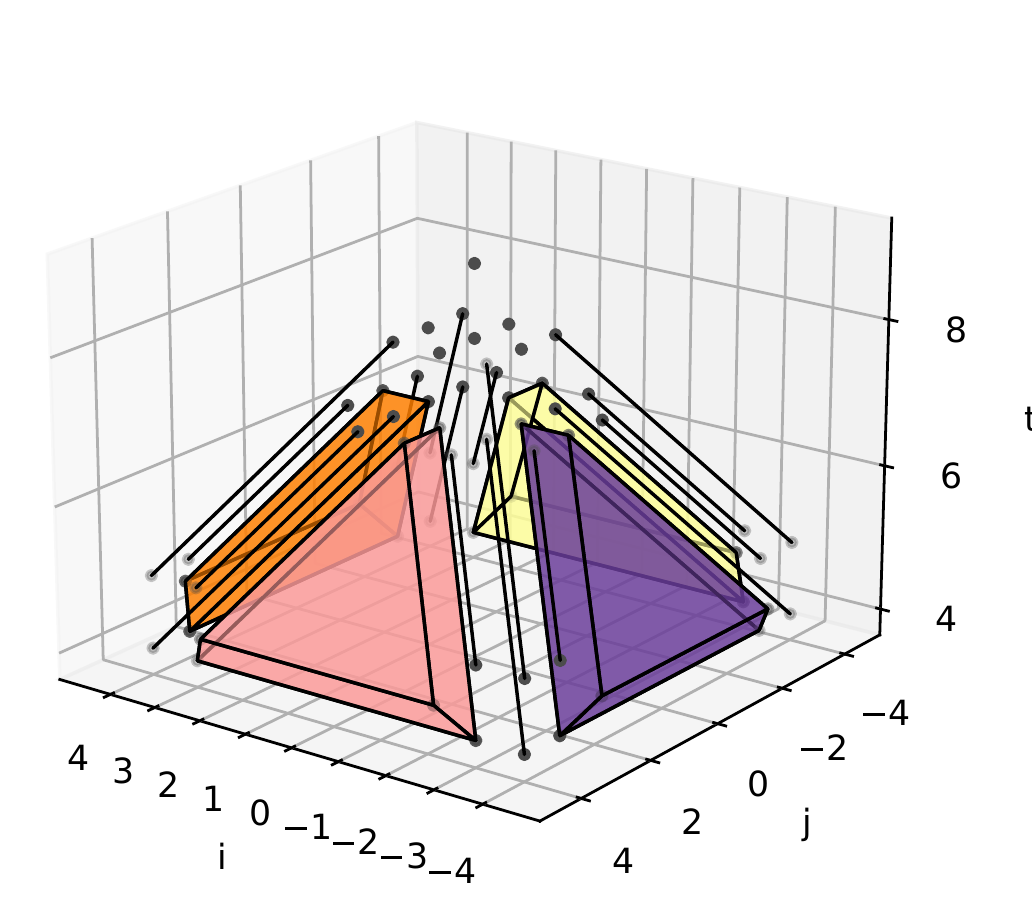}
    \caption{\texttt{jacobi-2d-d} (diamond tiling)}
    \label{fig:mars-jacobi-2d-diamond}
    \end{subfigure}
    \caption{Visualization of MARS generated using the MARS computer 
    (\url{https://github.com/cferr/mars.git}). Note that 
    $\vec{x}+\vec{b} \in \mathcal{D}$ isn't checked, the iteration space being
    assumed to be infinite by the computer.}
    \label{fig:mars-visualizations}
\end{figure*}

Table~\ref{tab:results} shows the results obtained by running the MARS computer
on the selected applications, and Figure~\ref{fig:mars-visualizations} shows
the MARS for a single tile of every application.

Table~\ref{tab:results} gives the number of dimensions of the iteration
space, the dependence pattern, tiling hyperplanes, the number of consumer tiles
per tile (\emph{\# Cons. Tiles}), the number of computed MARS (\emph{Nb MARS}) 
and the number of singleton MARS (that have a single point per tile). For the
\texttt{jacobi-2d-d} instance, there were 34 MARS computed, but only 26 of them
would not be empty when intersecting them with the tile of interest. The cause
of their emptiness has not yet been determined.

\subsubsection{Analysis}
Two observations can be made out of the MARS, on their number and the tiles that
consume them. As a general rule, the consumers tiles of a MARS are adjacent to 
each other, and the more cutting hyperplanes surrounding a MARS, the fewer 
dimensions it has. One notable case is \texttt{seidel-2d} 
(Figure~\ref{fig:mars-seidel-2d}) where a two-dimensional MARS is surrounded by 
two one-dimensional ones, close to the $t+i$ and $4t+2i+j$ hyperplanes 
intersection, and close to the $t$ and $4t+2i+j$ intersection.   

In 2-dimensional iteration spaces, two tiling hyperplanes are 
enough to tile all dimensions, in which case there are a maximum of 
$2^{2^2-1}-1 = 7$ MARS per tile. Our examples, \texttt{sw} and 
\texttt{jacobi-1d}, only exhibit 4 MARS, each MARS being consumed by two 
adjacent tiles. 

In 3-dimensional iteration spaces, the number of MARS goes up to 34 per tile
on \texttt{jacobi-2d-d} out of a maximum 32767 (due to the 15 consumer 
tiles). Computing the MARS took more than two hours for \texttt{jacobi-2d-d}
on an \emph{Intel Core i7-8665U} CPU.
As one can expect given the size of $\mathcal{P}_n(\mathcal{P}_n(\mathcal{H}))$, 
the complexity of the computation is such that our computer will not find the 
MARS in a matter of hours if there are 5 or more tiling hyperplanes. This is a
strong call to prune the search space. 

There are several possible optimizations. The first one is to figure out the
actual consumer tiles instead of enumerating 
$\mathcal{P}_n(\mathcal{P}_n(\mathcal{H}))$. This is implemented in the MARS
computer. 

Then, not all tuples of consumer tiles will yield a MARS; in particular, very 
specific dependence patterns will yield MARS consumed by non-adjacent tiles (for
instance, this is the case in \texttt{canonical-3d}, and more generally in any 
application where every dependence is orthogonal to a tiling hyperplane). 
Figuring out precisely when this happens would drastically reduce the search 
space. 

\section{Possible Applications}
\label{sec:applications}

MARS can be used in a variety of applications where fine-grain knowledge of the 
tile's flow-in origin and flow-out destination is known. In this paper, we 
detail two applications: compression, and memory allocation.

\subsection{Compression}
The fact that MARS are not redundant makes them suitable for compression: 
in general, decompression of an entire block of data is needed to access part of
it. When using MARS, all the data that is decompressed is actually needed, and
therefore there is no compression-induced redundancy. Works such as Ozturk et
al. \cite{Ozturk_2009} could be extended with data tiles of different sizes, where
each data tile is actually a MARS. 

Singleton and low-dimensional MARS are, however, going to be detrimental to 
compression. The volume of each MARS is a function of at least one tile size, 
with the exception of singleton MARS which volume will not change as tile size 
grows.

\subsection{Memory Allocation}

MARS can be used to construct a memory allocation for inter-tile communication,
similarly to what Zhao et al. \cite{Zhao_2021} have done with coarser-grain 
blocks. The idea is similar: allocate contiguous blocks of memory for each
MARS, and find a suitable layout.

\subsubsection{A case for merging MARS}
In some cases, as it can be observed in Figure~\ref{fig:mars-visualizations}, 
the no-redundancy property yielding singletons may cause performance penalties. 
This is notably the case when creating access functions at the granularity of 
MARS: such access functions will read or write the exact access data for each 
tile, but unless singleton accesses are merged with other accesses, these will 
incur a bandwidth waste.

To alleviate the performance issues, we need to relax the no-redundancy
property, and allow for MARS to be merged according to an objective
function. This merge process yields an intermediate partitioning between 
very fine-grain MARS and the entire flow-out, which would be the result of merging 
all MARS together.

\subsubsection{Global memory allocation}
The case of global memory allocation uses the number of transactions as a 
metric: the runtime of memory accesses is a function of the number of 
transactions and the volume of each transaction. Because global memories are
behind shared buses, each transaction costs a fixed initiation penalty caused
by arbitration, plus the number of cycles the actual transfer takes; the
longer the transfer, the more profitable it is and the better usage of the 
bandwidth. 

An optimization problem can be formulated with the following elements:
\begin{itemize}
  \item Minimize the number of transactions,
  \item Maximize every transaction's length (and therefore bandwidth usage),
  \item Find an intra-MARS layout,
  \item Find an inter-MARS layout, possibly allowing interleaving MARS from
  different tiles.
\end{itemize}

This is left for future work.

\section{Conclusion}
\label{sec:conclusion}

In this work, we have introduced an element of program analysis, MARS, to 
determine sets of data communicated between tiles without redundancy. These 
sets can be computed for certain programs with uniform dependence patterns,
and their computation can be automated.

A number of questions remain open with respect to the use of these sets. The
MARS are a fine-grain data structure in terms of usage per tile, which means
that they can yield the minumum amount of inter-tile communcation; however,
the presence of singletons with high transfer cost means that further
manipulations on MARS are needed to make them suitable for memory transfers or 
compression.

\bibliographystyle{ACM-Reference-Format}
\bibliography{refs}


\begin{thebibliography}{19}

%%% the Web address.

\ifx \showCODEN    \undefined \def \showCODEN     #1{\unskip}     \fi
\ifx \showDOI      \undefined \def \showDOI       #1{#1}\fi
\ifx \showISBNx    \undefined \def \showISBNx     #1{\unskip}     \fi
\ifx \showISBNxiii \undefined \def \showISBNxiii  #1{\unskip}     \fi
\ifx \showISSN     \undefined \def \showISSN      #1{\unskip}     \fi
\ifx \showLCCN     \undefined \def \showLCCN      #1{\unskip}     \fi
\ifx \shownote     \undefined \def \shownote      #1{#1}          \fi
\ifx \showarticletitle \undefined \def \showarticletitle #1{#1}   \fi
\ifx \showURL      \undefined \def \showURL       {\relax}        \fi
% invisible to TeX
\providecommand\bibfield[2]{#2}
\providecommand\bibinfo[2]{#2}
\providecommand\natexlab[1]{#1}
\providecommand\showeprint[2][]{arXiv:#2}

\bibitem[Bondhugula(2013)]%
        {Bondhugula_2013}
\bibfield{author}{\bibinfo{person}{Uday Bondhugula}.}
  \bibinfo{year}{2013}\natexlab{}.
\newblock \showarticletitle{Compiling Affine Loop Nests for Distributed-Memory
  Parallel Architectures}. In \bibinfo{booktitle}{\emph{Proceedings of the
  International Conference on High Performance Computing, Networking, Storage
  and Analysis}}. \bibinfo{publisher}{{ACM}}.
\newblock
\urldef\tempurl%
\url{https://doi.org/10.1145/2503210.2503289}
\showDOI{\tempurl}


\bibitem[Bondhugula et~al\mbox{.}(2017)]%
        {Bondhugula_2017}
\bibfield{author}{\bibinfo{person}{Uday Bondhugula}, \bibinfo{person}{Vinayaka
  Bandishti}, {and} \bibinfo{person}{Irshad Pananilath}.}
  \bibinfo{year}{2017}\natexlab{}.
\newblock \showarticletitle{Diamond Tiling: Tiling Techniques to Maximize
  Parallelism for Stencil Computations}.
\newblock \bibinfo{journal}{\emph{{IEEE} Transactions on Parallel and
  Distributed Systems}} \bibinfo{volume}{28}, \bibinfo{number}{5}
  (\bibinfo{date}{may} \bibinfo{year}{2017}), \bibinfo{pages}{1285--1298}.
\newblock
\urldef\tempurl%
\url{https://doi.org/10.1109/tpds.2016.2615094}
\showDOI{\tempurl}


\bibitem[Bondhugula et~al\mbox{.}(2008a)]%
        {Bondhugula_2008_CC}
\bibfield{author}{\bibinfo{person}{Uday Bondhugula}, \bibinfo{person}{Muthu
  Baskaran}, \bibinfo{person}{Sriram Krishnamoorthy}, \bibinfo{person}{J.
  Ramanujam}, \bibinfo{person}{A. Rountev}, {and} \bibinfo{person}{P.
  Sadayappan}.} \bibinfo{year}{2008}\natexlab{a}.
\newblock \showarticletitle{Automatic Transformations for
  Communication-Minimized Parallelization and Locality Optimization in the
  Polyhedral Model}. In \bibinfo{booktitle}{\emph{International Conference on
  Compiler Construction (ETAPS CC)}}.
\newblock
\urldef\tempurl%
\url{https://doi.org/10.1007/978-3-540-78791-4_9}
\showDOI{\tempurl}


\bibitem[Bondhugula et~al\mbox{.}(2008b)]%
        {Bondhugula_2008_PLDI}
\bibfield{author}{\bibinfo{person}{Uday Bondhugula}, \bibinfo{person}{Albert
  Hartono}, \bibinfo{person}{J. Ramanujam}, {and} \bibinfo{person}{P.
  Sadayappan}.} \bibinfo{year}{2008}\natexlab{b}.
\newblock \showarticletitle{A Practical Automatic Polyhedral Program
  Optimization System}. In \bibinfo{booktitle}{\emph{ACM SIGPLAN Conference on
  Programming Language Design and Implementation (PLDI)}}.
\newblock
\urldef\tempurl%
\url{https://doi.org/10.1145/1379022.1375595}
\showDOI{\tempurl}


\bibitem[Dathathri et~al\mbox{.}(2013)]%
        {Dathathri_2013}
\bibfield{author}{\bibinfo{person}{Roshan Dathathri}, \bibinfo{person}{Chandan
  Reddy}, \bibinfo{person}{Thejas Ramashekar}, {and} \bibinfo{person}{Uday
  Bondhugula}.} \bibinfo{year}{2013}\natexlab{}.
\newblock \showarticletitle{Generating Efficient Data Movement Code for
  Heterogeneous Architectures with Distributed-Memory}. In
  \bibinfo{booktitle}{\emph{Proceedings of the 22nd International Conference on
  Parallel Architectures and Compilation Techniques}}.
  \bibinfo{publisher}{{IEEE}}.
\newblock
\urldef\tempurl%
\url{https://doi.org/10.1109/PACT.2013.6618833}
\showDOI{\tempurl}


\bibitem[Feautrier(1991)]%
        {Feautrier_1991}
\bibfield{author}{\bibinfo{person}{Paul Feautrier}.}
  \bibinfo{year}{1991}\natexlab{}.
\newblock \showarticletitle{Dataflow analysis of array and scalar references}.
\newblock \bibinfo{journal}{\emph{International Journal of Parallel
  Programming}} \bibinfo{volume}{20}, \bibinfo{number}{1} (\bibinfo{date}{feb}
  \bibinfo{year}{1991}), \bibinfo{pages}{23--53}.
\newblock
\urldef\tempurl%
\url{https://doi.org/10.1007/BF01407931}
\showDOI{\tempurl}


\bibitem[Feautrier(1992a)]%
        {feautrier92a}
\bibfield{author}{\bibinfo{person}{Paul Feautrier}.}
  \bibinfo{year}{1992}\natexlab{a}.
\newblock \showarticletitle{Some Efficient Solutions to the Affine Scheduling
  Problem. {Part I}. One-dimensional Time}.
\newblock \bibinfo{journal}{\emph{International Journal of Parallel
  Programming}} \bibinfo{volume}{21}, \bibinfo{number}{5}
  (\bibinfo{year}{1992}), \bibinfo{pages}{313--347}.
\newblock
\urldef\tempurl%
\url{https://doi.org/10.1007/BF01407835}
\showURL{%
\tempurl}


\bibitem[Feautrier(1992b)]%
        {feautrier92b}
\bibfield{author}{\bibinfo{person}{Paul Feautrier}.}
  \bibinfo{year}{1992}\natexlab{b}.
\newblock \showarticletitle{Some Efficient Solutions to the Affine Scheduling
  Problem. {Part II}. Multidimensional Time}.
\newblock \bibinfo{journal}{\emph{International Journal of Parallel
  Programming}} \bibinfo{volume}{21}, \bibinfo{number}{6}
  (\bibinfo{year}{1992}), \bibinfo{pages}{389--420}.
\newblock
\urldef\tempurl%
\url{https://doi.org/10.1007/BF01379404}
\showURL{%
\tempurl}


\bibitem[Irigoin and Triolet(1988)]%
        {Irigoin_1988}
\bibfield{author}{\bibinfo{person}{F. Irigoin} {and} \bibinfo{person}{R.
  Triolet}.} \bibinfo{year}{1988}\natexlab{}.
\newblock \showarticletitle{Supernode partitioning}. In
  \bibinfo{booktitle}{\emph{Proceedings of the 15th {ACM} {SIGPLAN}-{SIGACT}
  symposium on Principles of programming languages - {POPL}
  {\textquotesingle}88}}. \bibinfo{publisher}{{ACM} Press}.
\newblock
\urldef\tempurl%
\url{https://doi.org/10.1145/73560.73588}
\showDOI{\tempurl}


\bibitem[Ozturk et~al\mbox{.}(2009)]%
        {Ozturk_2009}
\bibfield{author}{\bibinfo{person}{O. Ozturk}, \bibinfo{person}{M. Kandemir},
  {and} \bibinfo{person}{M.J. Irwin}.} \bibinfo{year}{2009}\natexlab{}.
\newblock \showarticletitle{Using Data Compression for Increasing Memory System
  Utilization}.
\newblock \bibinfo{journal}{\emph{{IEEE} Transactions on Computer-Aided Design
  of Integrated Circuits and Systems}} \bibinfo{volume}{28},
  \bibinfo{number}{6} (\bibinfo{date}{jun} \bibinfo{year}{2009}),
  \bibinfo{pages}{901--914}.
\newblock
\urldef\tempurl%
\url{https://doi.org/10.1109/tcad.2009.2017430}
\showDOI{\tempurl}


\bibitem[Pouchet and Yuki(2016)]%
        {Polybench}
\bibfield{author}{\bibinfo{person}{Louis-No{\"e}l Pouchet} {and}
  \bibinfo{person}{Tomofumi Yuki}.} \bibinfo{year}{2016}\natexlab{}.
\newblock \bibinfo{title}{PolyBench/C 4.2.1}.
\newblock
\newblock
\urldef\tempurl%
\url{http://polybench.sf.net}
\showURL{%
\tempurl}


\bibitem[Quinton and {Van Dongen}(1989)]%
        {quinton-jvsp89}
\bibfield{author}{\bibinfo{person}{P. Quinton} {and} \bibinfo{person}{V. {Van
  Dongen}}.} \bibinfo{year}{1989}\natexlab{}.
\newblock \showarticletitle{The Mapping of Linear Recurrence Equations on
  Regular Arrays}.
\newblock \bibinfo{journal}{\emph{Journal of {VLSI} Signal Processing}}
  \bibinfo{volume}{1}, \bibinfo{number}{2} (\bibinfo{year}{1989}),
  \bibinfo{pages}{95--113}.
\newblock
\urldef\tempurl%
\url{https://doi.org/10.1007/BF02477176}
\showURL{%
\tempurl}


\bibitem[Rajopadhye(1986)]%
        {sanjay-thesis}
\bibfield{author}{\bibinfo{person}{S.~V. Rajopadhye}.}
  \bibinfo{year}{1986}\natexlab{}.
\newblock \emph{\bibinfo{title}{Synthesis, Optimization and Verification of
  Systolic Architectures}}.
\newblock \bibinfo{thesistype}{Ph.\,D. Dissertation}.
  \bibinfo{school}{University of Utah}, \bibinfo{address}{Salt Lake City, Utah
  84112}.
\newblock


\bibitem[Rajopadhye(1989)]%
        {sanjay-dc}
\bibfield{author}{\bibinfo{person}{S.~V. Rajopadhye}.}
  \bibinfo{year}{1989}\natexlab{}.
\newblock \showarticletitle{Synthesizing Systolic Arrays with Control Signals
  from Recurrence Equations}.
\newblock \bibinfo{journal}{\emph{Distributed Computing}}  \bibinfo{volume}{3}
  (\bibinfo{date}{May} \bibinfo{year}{1989}), \bibinfo{pages}{88--105}.
\newblock
\urldef\tempurl%
\url{https://doi.org/10.1007/BF01558666}
\showURL{%
\tempurl}


\bibitem[Rajopadhye et~al\mbox{.}(1986)]%
        {sanjay-fst-tcs}
\bibfield{author}{\bibinfo{person}{S.~V. Rajopadhye}, \bibinfo{person}{S.
  Purushothaman}, {and} \bibinfo{person}{R.~M. Fujimoto}.}
  \bibinfo{year}{1986}\natexlab{}.
\newblock \showarticletitle{On Synthesizing Systolic Arrays from Recurrence
  Equations with Linear Dependencies}. In
  \bibinfo{booktitle}{\emph{Proceedings, Sixth Conference on Foundations of
  Software Technology and Theoretical Computer Science}}.
  \bibinfo{publisher}{Springer Verlag, LNCS~241}, \bibinfo{address}{New Delhi,
  India}, \bibinfo{pages}{488--503}.
\newblock
\urldef\tempurl%
\url{https://doi.org/10.1007/3-540-17179-7_30}
\showURL{%
\tempurl}


\bibitem[Renganarayanan et~al\mbox{.}(2012)]%
        {sanjay-toplas2012}
\bibfield{author}{\bibinfo{person}{Lakshminarayanan Renganarayanan},
  \bibinfo{person}{Daegon Kim}, \bibinfo{person}{Michelle~Mills Strout}, {and}
  \bibinfo{person}{Sanjay Rajopadhye}.} \bibinfo{year}{2012}\natexlab{}.
\newblock \showarticletitle{Parameterized loop tiling}.
\newblock \bibinfo{journal}{\emph{ACM Transactions on Programming Languages and
  Systems (TOPLAS)}} \bibinfo{volume}{34}, \bibinfo{number}{1}
  (\bibinfo{year}{2012}), \bibinfo{pages}{3}.
\newblock
\urldef\tempurl%
\url{https://doi.org/10.1145/2160910.2160912}
\showDOI{\tempurl}


\bibitem[Verdoolaege and Grosser(2012)]%
        {Verdoolaege_2012}
\bibfield{author}{\bibinfo{person}{Sven Verdoolaege} {and}
  \bibinfo{person}{Tobias Grosser}.} \bibinfo{year}{2012}\natexlab{}.
\newblock \showarticletitle{Polyhedral Extraction Tool}. In
  \bibinfo{booktitle}{\emph{Second Int. Workshop on Polyhedral Compilation
  Techniques (IMPACT'12)}}. \bibinfo{address}{Paris, France}.
\newblock


\bibitem[Wolf and Lam(1991)]%
        {Wolf_1991}
\bibfield{author}{\bibinfo{person}{Michael~E. Wolf} {and}
  \bibinfo{person}{Monica~S. Lam}.} \bibinfo{year}{1991}\natexlab{}.
\newblock \showarticletitle{A data locality optimizing algorithm}.
\newblock \bibinfo{journal}{\emph{{ACM} {SIGPLAN} Notices}}
  \bibinfo{volume}{26}, \bibinfo{number}{6} (\bibinfo{date}{jun}
  \bibinfo{year}{1991}), \bibinfo{pages}{30--44}.
\newblock
\urldef\tempurl%
\url{https://doi.org/10.1145/113446.113449}
\showDOI{\tempurl}


\bibitem[Zhao et~al\mbox{.}(2021)]%
        {Zhao_2021}
\bibfield{author}{\bibinfo{person}{Tuowen Zhao}, \bibinfo{person}{Mary Hall},
  \bibinfo{person}{Hans Johansen}, {and} \bibinfo{person}{Samuel Williams}.}
  \bibinfo{year}{2021}\natexlab{}.
\newblock \showarticletitle{Improving communication by optimizing on-node data
  movement with data layout}. In \bibinfo{booktitle}{\emph{Proceedings of the
  26th {ACM} {SIGPLAN} Symposium on Principles and Practice of Parallel
  Programming}}. \bibinfo{publisher}{{ACM}}.
\newblock
\urldef\tempurl%
\url{https://doi.org/10.1145/3437801.3441598}
\showDOI{\tempurl}


\end{thebibliography}

\section*{Acknowledgments}
The authors would like to thank Tomofumi Yuki for the scientific insight and 
valuable discussions that led to the construction of MARS.

\end{document}